\documentclass[apj]{emulateapj}
\usepackage{mathptmx}

\def\gtorder{\mathrel{\raise.3ex\hbox{$>$}\mkern-14mu
             \lower0.6ex\hbox{$\sim$}}}
\def\ltorder{\mathrel{\raise.3ex\hbox{$<$}\mkern-14mu
             \lower0.6ex\hbox{$\sim$}}}

\slugcomment{Draft of \today}

\shorttitle{GRB~070201}

\shortauthors{Ofek et al.}

\begin{document}

\title{GRB~070201: A possible Soft Gamma Ray Repeater in M31\footnote{Based on observations obtained with {\it XMM}-Newton, an ESA science mission with 
instruments and contributions directly funded by ESA Member States and NASA}}
\author{
E.~O.~Ofek\altaffilmark{1},
M.~Muno\altaffilmark{1},
R.~Quimby\altaffilmark{1},
S.~R.~Kulkarni\altaffilmark{1},
H.~Stiele\altaffilmark{3},
W.~Pietsch\altaffilmark{3},
E.~Nakar\altaffilmark{1},
A.~Gal-Yam\altaffilmark{4},
A.~Rau\altaffilmark{1},
P.~B.~Cameron\altaffilmark{1},
S.~B.~Cenko\altaffilmark{1},
M.~M.~Kasliwal\altaffilmark{1},
D.~B.~Fox\altaffilmark{2},
P.~Chandra\altaffilmark{5}$^{,}$\altaffilmark{6}
A.~K.~H.~Kong\altaffilmark{7}$^{,}$\altaffilmark{8}
and R.~Barnard\altaffilmark{9}
}
\altaffiltext{1}{Division of Physics, Mathematics and Astronomy, California Institute of Technology, Pasadena, California 91125, USA}
\altaffiltext{2}{Department of Astronomy and Astrophysics, 525 Davey Laboratory, Pennsylvania State University, University Park, Pennsylvania 16802, USA}
\altaffiltext{3}{Max-Planck-Institut fur extraterrestrische Physik, Giessenbachstr. 1, D-85740 Garching, Germany}
\altaffiltext{4}{Physics Faculty, Weizmann Institute of Science, Rehovot 76100, Israel}
\altaffiltext{5}{Jansky Fellow, National Radio Astronomical Observatory}
\altaffiltext{6}{Department of Astronomy, University of Virginia, P.O.Box 400325, Charlottesville, VA 22904}
\altaffiltext{7}{Kavli Institute for Astrophysics and Space Research, Massachusetts Institute of Technology, 77 Massachusetts Avenue, Cambridge, MA 02139}
\altaffiltext{8}{Department of Physics and Institute of Astronomy, National Tsing Hua University, Hsinchu, Taiwan}
\altaffiltext{9}{The Open University, Walton Hall, Milton Keynes, MK7 6AA, UK}

\begin{abstract}

The gamma-ray burst (GRB) 070201 was a bright short-duration
hard-spectrum
GRB detected by the Inter-Planetary Network (IPN).
Its error quadrilateral, which has an area of $0.124$~deg$^{2}$,
intersects some prominent spiral arms of the nearby M31 (Andromeda) galaxy.
Given the properties of this GRB, along with the fact
that LIGO data argues against a compact binary merger origin
in M31, this GRB is an excellent candidate for an
extragalactic Soft Gamma-ray Repeater (SGR) giant flare,
with energy of $1.4\times10^{45}$~erg.
Analysis of ROTSE-IIIb visible light observations of M31,
taken $10.6$~hours after the burst
and covering 42\% of the GRB error region,
did not reveal any optical transient
down to a limiting magnitude of 17.1.
We inspected archival and proprietary
{\it XMM}-Newton X-ray observations of the intersection of
the GRB error quadrilateral and M31,
obtained about four weeks prior to the outburst, in order
to look for periodic variable X-ray sources.
No SGR or Anomalous X-ray Pulsar (AXP) candidates
(periods in range 1 to 20~s) were detected.
We discuss the possibility of detecting extragalactic SGRs/AXPs
by identifying their periodic X-ray light curves.
Our simulations suggest that the probability of detecting
the periodic X-ray signal of
one of the known Galactic SGRs/AXPs,
if placed in M31,
is about $10\%$ ($50\%$),
using 50~ks (2~Ms) {\it XMM}-Newton exposures.

\end{abstract}


\section{Introduction}
\label{Introduction}

Soon after the discovery of Gamma-Ray Bursts (GRBs),
it was realized that some bursts are repeating.
The localization of these objects, called
Soft Gamma-Ray Repeaters (SGRs), showed that they
are located in the local group
(Cline et al. 1980; Evans et al. 1980; Mazets \& Golenetskii 1981),
and their flare energy release ranges from
$\sim10^{39}$ to $10^{46}$~erg.

In quiescence, SGRs (and also the related
class of Anomalous X-ray Pulsars; AXPs)
are detected as faint X-ray sources
with luminosities in the range $\sim10^{33}$ to $10^{36}$~erg~s$^{-1}$.
Their X-ray light curves
are modulated with periodicities of the order of 10~s,
and period derivatives of the order of $10^{-10}$~s~s$^{-1}$.
These properties suggest that SGRs are young neutron stars
with ultra-strong magnetic fields ($\gtorder10^{14}$~G).
Contrary to ``normal'' neutron stars (i.e., radio pulsars),
whose energy reservoir is rotational,
SGR's source of energy is most probably magnetic.
The basic properties of SGRs are well explained by the popular
magnetar model (Duncan \& Thompson 1992; Paczynski 1992).

Known SGRs and AXPs are associated with star forming regions
(for a review see Gaensler et al. 2001).
Moreover, some of them may be associated
with supernova remnants
(Cline et al. 1982; Kulkarni \& Lingenfelter 1994;
 Hurley et al. 1999;
 Woods et al. 1999).
However, Levan et al. (2006) suggested a formation
channel for magnetars in old stellar populations.
%

Unfortunately, only four SGRs are known to date, all in the
local group (see Woods \& Thompson 2006 for a recent review),
of which three reside in the Milky-Way and one in the
Large Magellanic Cloud.
The small number of known SGRs severely hinders our ability to study
their origin, 
environments (e.g., Gaensler et al. 2001),
and rate of luminous flares
(Palmer et al. 2005; Popov \& Stern 2006; Ofek 2007a).

However, the strongest SGR flares can be detected in nearby
galaxies (e.g., Duncan 2001; Eichler 2002).
Discovery of extragalactic SGRs is
an exciting possibility that will
enable us to enlarge the sample
of known objects in this class.
Unfortunately, extragalactic SGR flares have proven
hard to recognize, and their observed rate
of gamma-ray flares
is probably of the order of several percent
of the short-duration GRB observed rate
(e.g., Lazzati et al. 2005; Nakar et al. 2006; Ofek 2007a).
To date, only a small number of extragalactic SGR candidates
are known:
in M81 (Ofek et al. 2006; Frederiks et al. 2007b), in NGC\,6946 (Crider 2006),
and in M74 (Ofek 2007a).
Unfortunately, each of these candidates has been observed to flare only once.
Moreover, because of the limited positional accuracy
of most current gamma-ray telescopes,
they have astrometric uncertainties of
hundreds of square arcminutes or more.
This positional accuracy is too poor to allow environmental studies.
Furthermore,
given the relatively large positional uncertainty,
it is possible that some of these
candidates are due to a chance coincidence.

Discovery of extragalactic SGRs will increase our statistical sample
of such objects and
with accurate positions it will be possible to study
their environments.
In particular, it may reveal a new population of
SGRs that are not bound to star forming
regions (e.g., Levan et al. 2006).

\subsection{GRB~070201}
\label{GRB070201}

In this paper we discuss an extragalactic
SGR giant flare candidate associated with the nearby galaxy M31.
At UTC 2007 Feb 1, 15:23:10.780, an intense short-hard GRB
with $\sim0.2\,$s duration was detected by the
Inter-Planetary Network (IPN; e.g. Hurley et al. 1999).
The burst was detected by Konus-{\it Wind},
{\it INTEGRAL}-SPI-ACS,
{\it Swift}-BAT\footnote{The burst was outside the BAT coded field of view. Therefore it was~not localized by {\it Swift}-BAT.}
(Golenetskii et al. 2007a),
and {\it Messenger},
while {\it Suzaku} and {\it RHESSI} were not able to observe the burst due to
Earth occultation, and {\it Odyssey} was not able to observe
it due to Mars occultation.
Early on, Perley \& Bloom (2007) noticed that the
preliminary IPN annulus crosses the Andromeda galaxy
(see also: Ofek 2007b; Golenetskii et al. 2007b).
Later on, with the analysis of the {\it Messenger}
data (Hurley et al. 2007), and re-analysis
of the data (Mazets et al. 2007),
the error region shrunk to a $0.124$~deg$^{2}$
quadrilateral that intersects the M31 galaxy.
\begin{figure}
\centerline{\includegraphics[width=8.5cm]{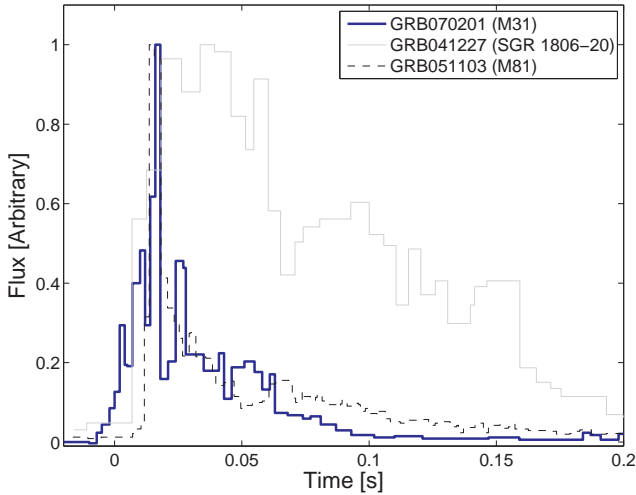}}
\caption{The Konus-{\it Wind} gamma-ray light curve of GRB\,070201 (solid heavy line),
compared with the light curve of the 2004 December 27 SGR giant flare
(dotted line), and the Konus-{\it Wind} light curve of GRB\,051103
(dashed-dotted line; Ofek et al. 2006; Frederiks et al. 2007b).
The light curve of the 2004 December 27 SGR giant flare is based
on a digitization of Figure~1 in Terasawa et al. (2005),
while the light curve of GRB~070201 is based on a digitization of
the 18-1160 keV-band light curve from the Konus-{\it Wind} website.
The fluxes of the different bursts are scaled such they will have the same
peak flux.
\label{GRB070201_LC}}
\end{figure}
\begin{figure*}
\centerline{\includegraphics[width=18cm]{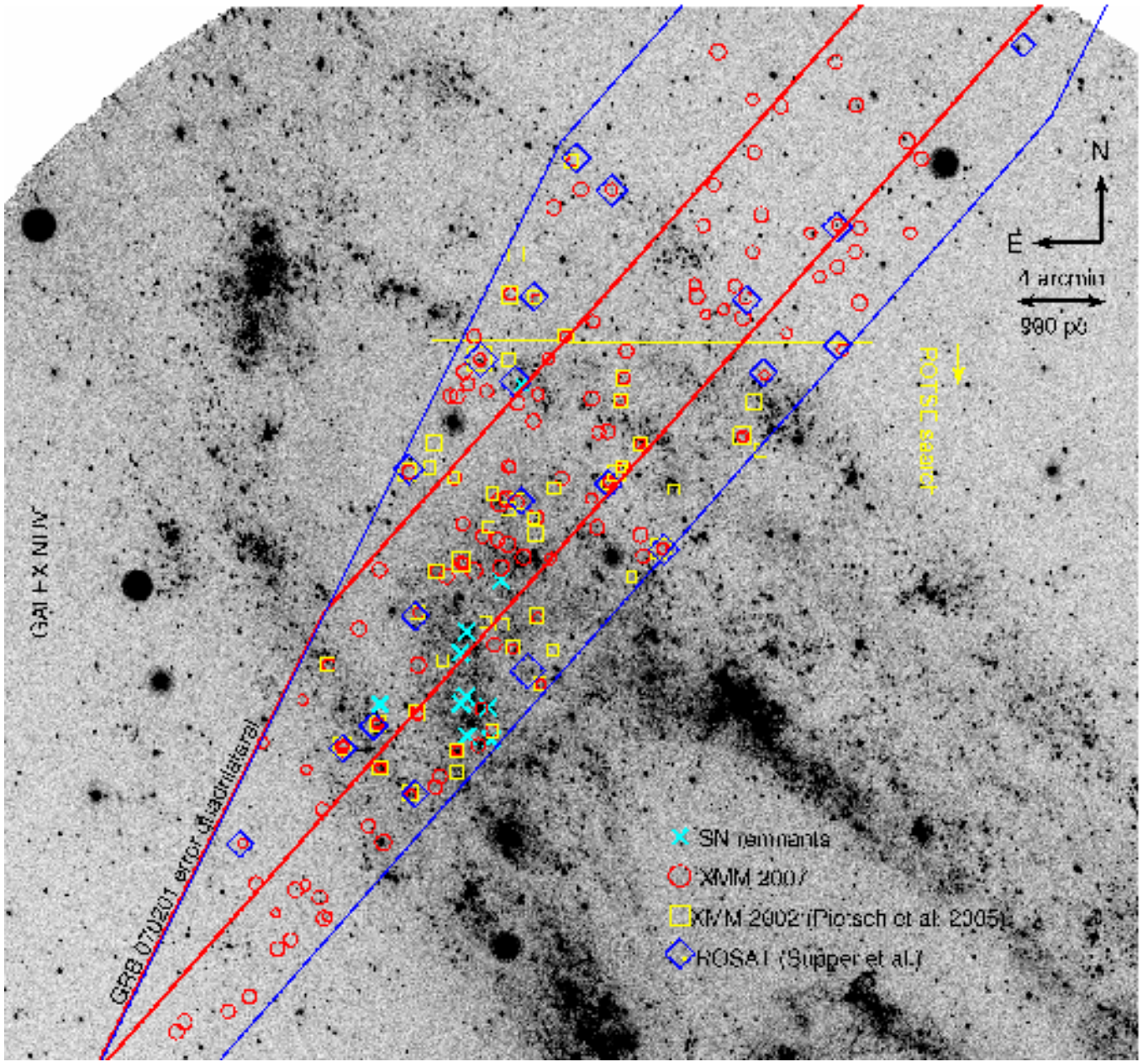}}
\caption{{\it GALEX} Near-UV image,
obtained on UTC 2003 September 5 (1940~s exposure),
of the region of the error quadrilateral
of GRB\,070201 intersecting
with M31. The solid red lines mark the
revised error quadrilateral (Mazets et al. 2007)
and the blue lines show the original error quadrilateral
(Hurley et al. 2007),
the red circles show the position of X-ray sources
detected during the 2007 {\it XMM} observations, the yellow boxes
mark the {\it XMM} X-ray sources detected in 2002,
the blue diamonds mark the position of X-ray sources
listed in the {\it ROSAT}-PSPC catalog of M31 (Supper et al. 2001), and the
the cyan crosses mark the position of known supernova (SN) remnants in M31
(Magnier et al. 1995).
The size of the markers (in arcsec) of the X-ray sources corresponds to
their flux ($F$) in the 0.2--10~keV band using the relation:
$10+10\times(15+log_{10}{F [{\rm erg}~{\rm s}^{-1}~{\rm cm}^{-2}]})$.
More than one symbol of the same type at
almost the same position corresponds to detection
of the same source in the overlap regions between images
taken during the same year.
The ROTSE-IIIb observations cover the entire error region south of
the yellow line.
\label{GALEX_X}}
\end{figure*}

The burst had the highest peak count rate of any
GRB observed by Konus-{\it Wind} in 12 years of operation
(excluding Galactic SGRs).
The GRB fluence in the Konus-{\it Wind} $20\,$keV--$1.2\,$MeV band was
$2.00_{-0.26}^{+0.10}\times10^{-5}\,$erg\,cm$^{-2}$,
and its peak flux on two-milliseconds time scale was
$1.61_{-0.50}^{+0.29}\times10^{-3}\,$erg\,cm$^{-2}$\,s$^{-1}$
($90\%$~confidence; Mazets et al. 2007).
The light curve, shown in Figure~\ref{GRB070201_LC} (solid line; Mazets et al. 2007),
had a ``bumpy'' rise with a time scale of $20$~milliseconds
and two leading peaks with durations
of a few milliseconds, while
the decaying tail had a time scale of about $0.1\,$s
(see discussion in Mazets et al. 2007).

%
Golenetskii et al. (2007b) and Mazets et al, (2007) found that
the spectrum of GRB\,070201 is well fitted by a power-law
with an exponential cutoff,
$dN/dE \sim E^{-\alpha}\exp{[-E(2-\alpha)/E_{p}]}$,
where $E$ is the energy.
They also found that,
in the first $64\,$milliseconds,
the best fit parameters
are $\alpha=0.52_{-0.15}^{+0.13}$, and $E_{p}=360_{-38}^{+44}\,$keV
($90\%$ confidence; $\chi^{2}/dof=32/35$),
while the best fit parameters for the time integrated spectrum
are $\alpha=0.98_{-0.11}^{+0.10}$, and $E_{p}=296_{-32}^{+38}\,$keV
($\chi^{2}/dof=40/40$).
Like GRB~070201,
the spectrum of the 2004 December 27
giant flare of SGR\,1806$-$20, at peak, is not consistent
with a black-body spectrum, but is well fitted by
a power-law with an exponential cutoff model,
with $\alpha=0.73_{-0.64}^{+0.47}$ ($\chi^{2}/dof=10.6/12$;
Frederiks et al. 2007a).

Abbott et al. (2007b) analyzed the available
Laser Interferometric Gravitational-Wave Observatory
(LIGO; Abbott et al. 2007a) data,
collected within 180~s of the time of GRB~070201.
They did~not find any gravitational wave source
coincident with this GRB.
Using these observations they rule
out, at the $99\%$ confidence level, a compact binary
(i.e., black-holes or neutron stars) merger
origin for this GRB with progenitors
masses in the ranges:
$1~{\rm M}_{\odot} < M_{1} < 3~{\rm}_{\odot}$ and
$1~{\rm M}_{\odot} < M_{2} < 40~{\rm}_{\odot}$,
and with a distance below 3.5~Mpc.

In this paper, we present the case of GRB\,070201 as a possible SGR
giant flare in the nearby galaxy M31.
In \S\ref{Obs} we present
our search for a visible light transient associated with
this GRB.
We examine archival X-ray and UV
images of the IPN error quadrilateral (\S\ref{Archive}),
and we look for possible candidates for
pulsating X-ray sources that could be SGRs within
M31 (\S\ref{Xray}). In \S\ref{Sim} we quantify
the probability to detect X-ray pulsations of
an SGR or AXP in M31, and finally we discuss
the nature of GRB~070201 in \S\ref{Disc}.

\section{Observations}
\label{Obs}

Optical images of the Andromeda galaxy were obtained nightly by the
0.45-m ROTSE-IIIb telescope as part of
the Texas Supernova Search (Quimby 2006).
Routine unfiltered images covering the GRB error quadrilateral
south of $\delta=+42^{h}08^{m}57^{s}$
(i.e., the southern $42\%$ of the error box, including the
intersection with the
spiral arms) were taken on
UTC 2007 Feb 02.0821, 10.6 hours after the GRB trigger.

We preformed PSF-matched image subtraction of
the data using a modified version of the
Supernova Cosmology Project's search code (Perlmutter et al. 1999).
After subtracting off a reference template constructed from 37
ROTSE-IIIb images obtained between 2005
July and 2006 June, we find no new objects in the southern part of
the error box covered to a 5-$\sigma$
limiting magnitude of 17.15 (calibrated against the USNO-B1.0 R2 magnitude).
Assuming a distance to M31 of 770~kpc (e.g., Ribas et al. 2005),
and correcting for Galactic extinction in this direction
(Schlegel, Finkbeiner, \& Davis 1998; Cardelli, Clayton, \& Mathis 1989),
this corresponds to an absolute
magnitude limit of $-7.4$.

\section{Archival data}
\label{Archive}

The intersection region
of the error quadrilateral of GRB\,070201 with M31
has been observed by several facilities, including
{\it ROSAT}, {\it GALEX} and {\it XMM}-Newton.
{\it XMM} observed this field on several epochs,
listed in Table~\ref{Tab-ObsLog}.
Analysis of the 2002 {\it XMM}-Newton data
was presented in Pietsch, Freyberg, \& Haberl (2005).
Interestingly, the last {\it XMM} observation of the field was obtained
about four weeks prior to the GRB trigger, as part of the 
M31 {\it XMM}-Newton X-ray survey (Stiele et al. 2007).
Source extraction from the 2007 {\it XMM} images
is presented in \S\ref{Xray}, while a complete
catalog and analysis of the 2007 {\it XMM} M31 observations
will be presented in Stiele et al. (2008, in preparation).
\begin{deluxetable}{ccccc}
\tablecolumns{5}
\tablewidth{0pt}
\tablecaption{Log of Newton-{\it XMM} observations}
\tablehead{
\colhead{Date}       &
\colhead{Exp.}       &
\colhead{R.A.}       &
\colhead{Dec.}       &
\colhead{PA}        \\
\colhead{}           &
\colhead{ks}         &
\colhead{deg}        &
\colhead{deg}        &
\colhead{deg}        
}
\startdata
%
 2002-01-26.7 &  4   & 11.33543 & $+$41.93236 & 237.28 \\
 2002-01-27.0 & 54   & 11.36929 & $+$41.92389 & 237.24 \\
 2007-01-02.9 & 54   & 11.46008 & $+$41.51242 & 251.68 \\
 2007-01-04.2 & 52   & 11.23317 & $+$42.14294 & 250.50 \\
 2007-01-04.9 & 62   & 11.69142 & $+$41.88250 & 250.71 \\
 2007-01-06.2 & 55   & 11.36483 & $+$41.91969 & 249.36 \\
\enddata
\tablecomments{{L}ist of {\it XMM}-Newton observations of the error
quadrilateral of GRB~070201. PA is the position angle of the
{\it XMM}-Newton instruments.}
\label{Tab-ObsLog}
\end{deluxetable}

In Figure~\ref{GALEX_X} we present the {\it GALEX} Near-UV image
of the region of the error quadrilateral intersecting
M31. In this Figure, we show: the
refined (red lines; Mazets et al. 2007)
and original (blue lines; Hurley et al. 2007)
IPN error quadrilateral;
the {\it ROSAT} PSPC sources (blue diamonds; Supper et al. 2001);
the {\it XMM} sources detected in 2002 (yellow boxes; Pietsch et al. 2005);
the {\it XMM} sources detected in 2007 (red circles);
and known and candidates SN remnants (Magnier et al. 1995).
The size of the markers of X-ray sources correspond to
their flux.
We note that there is some overlap between
the {\it XMM} observations.
Therefore, more than one symbol of the same type in
almost the same position corresponds to a detection
of the same source in the overlap regions between images
taken the same year.
The ROTSE-IIIb observations covers the error quadrilateral
south of the yellow line.

Several X-ray sources in the field of GRB\,070201
(Fig.~\ref{GALEX_X})
show long timescale variability
between 2002 and 2007.
Since several types of astrophysical X-ray sources
are known to vary,
this information by itself is not very constructive for
the identification of an SGR X-ray counterpart in this field.
However, an SGR or an AXP may reveal itself as a pulsating
X-ray source with periodicity around 10~s.
In the next section we describe a search for
such X-ray variable sources.
A thorough variability analysis of X-ray sources
in the entire M31 galaxy will be presented in
Stiele et al. (2008, in preparation).

\section{Search for short-period X-ray variable sources in the error quadrilateral}
\label{Xray}

All known SGRs/AXPs exhibit X-ray pulsations
with periodicities in the range of 2~s to 12~s.
Therefore,
it may be possible to identify such objects in M31
by looking for X-ray variable sources with periods
in the range of $\sim$1 to 20~s.

To look for such sources,
we inspected the pipeline-processed event files
of the four fields observed in 2007 (Table~\ref{Tab-ObsLog}),
and removed time intervals during which
particle events caused the event rate 
in the detector to flare by more than two standard deviations
above the mean rate.
We then created images
of the 0.2--12~keV events, binned to $4$~arcsec resolution .
For the purpose of identifying point sources, standard data
selection was applied to make the images,
to remove events near the edges of 
detector chips and bad pixels, and
to reject events that were likely to be cosmic rays (pattern $>4$
for the PN, and $>12$ for the MOS detectors).
We then generated matching exposure maps, and
we searched for point sources using the routine {\tt ewavelet},
separately for each detector.
We extracted events for each source from 
the radius defined by {\tt ewavelet}, which was
$\approx15$~arcsec.
This radius contains about $50\%$ of the photons for each source.
The arrival times of the photons
were transformed to the solar system barycenter using the tool {\tt barycen}.
Finally, we searched for periodicities
in the extracted time tagged photons
using discrete fast
Fourier transforms.
The time series were padded so that the number of points
in the transforms were a power of 2. 
This provides a frequency resolution slightly finer than
$1/t_{\rm exp}$, where $t_{\rm exp}$ is the exposure time.
The maximum frequency considered was the Nyquist frequency of
the 13.6~Hz PN detector sampling rate, and 
the lowest frequency searched was $10^{-4}$~Hz.
We found no signals stronger than 19.15 times
the mean of the power-spectrum noise.
This cutoff power was selected,
such the
probability of a
single source to surpass this threshold,
in one or more of the $\sim10^{6}$ tested frequencies,
is about $\sim1\%$ (in the entire FFT-tested frequencies range).
Limiting ourself to the 1~s to 20~s periodicity range,
this limit corresponds to false alarm probability
of $\sim0.05\%$.
In total we search for periodicity among 149
X-ray sources, within the original error quadrilateral
(Hurley et al. 2007; blue lines in Fig.~\ref{GALEX_X})
which are listed in Table~\ref{Tab-src}.
We did~not find any periodic variable among the {\it XMM} X-ray
sources.
\begin{deluxetable}{llllll}
\tablecolumns{6}
\tablewidth{0pt}
\tablecaption{{L}ist of X-ray sources for which periodicity was searched}
\tablehead{
\colhead{Name} &
\colhead{RA (J2000)} &
\colhead{Dec (J2000)} &
\colhead{r\tablenotemark{a}} &
\colhead{Counts} &
\colhead{Obs/Det\tablenotemark{b}} \\
\colhead{} &
\colhead{deg} &
\colhead{deg} &
\colhead{$''$} &
\colhead{} &
\colhead{}
}
\startdata
004603.5$+$414623 & 11.51444 & 41.77310 &  18.1 &  42274 &  201/PN \\
004617.7$+$414258 & 11.57362 & 41.71622 &  29.5 &  24500 &  201/PN \\
004618.7$+$414354 & 11.57812 & 41.73170 &  14.1 &  19633 &  201/PN \\
004624.6$+$414414 & 11.60240 & 41.73723 &  18.3 &  22009 &  201/PN \\
004625.6$+$414159 & 11.60687 & 41.69995 &  13.8 &   9300 &  201/PN \\
\enddata
\tablenotetext{a}{Aperture radius in which source counts were extracted.}
\tablenotetext{b}{{L}ast three digits of the {\it XMM} observation ID (starts with 0402561) followed by the detector name (M1, M2 or PN).}
\tablecomments{First five entries of the Table. The Table in its entirety is available via the electronic version.}
\label{Tab-src}
\end{deluxetable}
%
%
%
%

\section{Is it possible to detect the modulated X-ray emission of SGRs in M31?}
\label{Sim}

The quiescence X-ray luminosity of known AXPs/SGRs ranges from
$10^{33}$ to $10^{36}$~erg~s$^{-1}$.
At the distance of M31 ($770$~kpc, e.g., Ribas et al. 2005),
these correspond to fluxes
of $10^{-17}$ to $10^{-14}$~erg~s$^{-1}$~cm$^{-2}$
in the $2$--$10$~keV range.
Given these flux levels, we discuss here
the chances to detect the modulated X-ray light curves
of SGRs/AXPs in M31 as a function of
the flux of the X-ray source and its
light curve shape (i.e., the fraction of flux within
a pulse).
Specifically, we would like to answer the question:
what is the probability to detect an SGR or an AXP,
based on its periodic X-ray signal,
in the Andromeda galaxy? --
In order to answer this question we perform the
simulations described below.

In our simulations we assumed a 75~ks exposure
with the {\it XMM}-Newton fully depleted PN CCD,
which roughly corresponds to a 50~ks integration
with all the European Photon Imaging Camera (EPIC) CCDs.
Our simulated time-tagged X-ray light curves consist
of the background expected for an {\it XMM} observation
and a periodic signal.
The periodic light curve consists of a non variable part
and photons clumped in periodic pulses.
In all the simulated photon-tagged light curves
the periodicity was set to exactly 10~s,
and the width of the periodic pulse
was $20\%$ of the period (i.e., 2~s).
We controlled the ``shape'' of the light curve by adjusting
the fraction of photons
within a pulse (hereafter 'pulse fraction').

We simulated light curves in a dense grid 
of count rates and pulse fractions.
The counts rate were set to be between $10^{-4}$ to $5\times10^{-2}$
counts per sec (along 100 logarithmically spaced grid points), 
and the pulse fractions in the range
$0.21$ to $0.81$ (61 linearly spaced grid points).
In each grid point we simulated 100 photon-tagged
light curves, and for each light curve
we calculated the power spectrum,
and checked if the 10~s period
signal is stronger than 
19.15 times
the mean of the power-spectrum noise.
We note that this threshold was used in the search for
X-ray variable sources described in \S\ref{Xray}.
Finally, in each grid point
we calculated the probability to recover
the periodic signal with a power exceeding that threshold,
which corresponds to false alarm probability of
about $1\%$.

\begin{figure}
\centerline{\includegraphics[width=8.5cm]{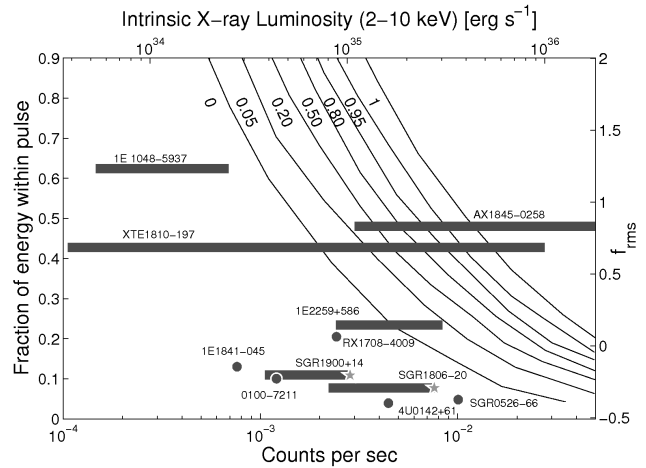}}
\caption{The probability (contours) to detect,
with $1\%$ false alarm probability,
each one of the Galactic AXPs/SGRs (in quiescent state)
if placed in the Andromeda
galaxy, as a function the source count rate
(lower X-axis) or luminosity (upper X-axis),
and the fraction of energy within
a $20\%$ (of period) width pulse (left Y-axis).
Translation of the energy within the pulse
to rms pulse fraction, $f_{rms}$
(for definition see Woods \& Thompson [2006; Table 14.2]), is shown
on the right-hand Y-axis.
The location of known AXPs/SGRs are shown
as circles or lines (if variable).
The simulations assume the {\it XMM}-PN is observing
the targets for 75~ks which is roughly
equivalent to the 50~ks {\it XMM} observations we analyzed.
The properties of the SGRs/AXPs
(i.e., luminosity range, spectral shape, and rms pulse fraction)
were adopted from Woods \& Thompson (2006).
To account for the observed elevated X-ray luminosity of SGRs
about one month prior to giant flares,
we increased the maximum quiescence luminosities
of SGR~1900$+$14 and SGR~1806$-$20 by 1.5 and 2, respectively
(Woods et al. 2001; 2007).
These elevated luminosities are marked as stars
in the Figure.
\label{AXP_DetSim}}
\end{figure}

Figure~\ref{AXP_DetSim} presents the result of these
simulations.
The contours show the probability to detect
the X-ray periodicity with a
false alarm probability of $1\%$ per each source
(assuming that in each source $10^{6}$ independent
frequencies are tested),
as a function of the two free parameters.
The lower X-axis shows the observed count rate
(and the luminosity at the distance of M31,
on the upper X-axis),
and the Y-axis marks the fraction of energy
within a pulse whose width is $20\%$ of the period
of the light curve.
On the right-hand Y-axis we show the rms pulsed fraction,
$f_{rms}$, defined in Woods \& Thompson (2006; Table 14.2).

Next we compared these simulations
with the actual
properties of known AXPs and SGRs.
For each one of the 11 AXPs and SGRs
listed in Woods \& Thompson (2006),
for which the luminosity and rms pulse fraction ($f_{rms}$)
are known, we calculated their count rates (or range
of count rates in case they are variable).
We converted the luminosity of the AXPs/SGRs to count rates using the
PIMMS web tool\footnote{http://cxc.harvard.edu/toolkit/pimms.jsp},
and assumed a neutral Hydrogen column density of $10^{21}$~cm$^{-2}$,
in the direction of M31 (Dickey \& Lockman 1990; Kalberla et al. 2005),
and that the distance to M31 is 770~kpc (e.g., Ribas et al. 2005).
Furthermore, we assumed that the X-ray spectrum of each SGR/AXP is described
only\footnote{Note that some of these objects have more complicated spectra.} by
a power-law and we adopted the measured power-law indices
for each one of these sources (Woods \& Thompson 2006).
The location of the known SGRs/AXPs in the pulse fraction
vs. X-ray luminosity (in the $2-10$~keV
range\footnote{The simulation assumes the observations are conducted
in the 0.2--10~keV band. For compatibility with
Woods \& Thompson (2006), we present the luminosity
in the $2-10$~keV band.}) space are presented in
Fig.~\ref{AXP_DetSim}
as circles (or lines to indicate a range).

The X-ray emission about one month
before and several months after an SGR giant flare
is known to be higher than ``normal''.
This may elevate the probability to detect
X-ray emission from extragalactic SGR giant flares
in the {\it XMM} M31 images taken four weeks prior to the burst.
For example, the X-ray flux of SGR~1900$+$14 was about
1.5 times higher than normal, starting about one month prior
to the SGR giant flare of 1998 Aug 27,
and also for a year past the flare.
In the case of the
December 27, 2004 giant flare,
the X-ray emission from SGR~1806$-$20 
was about two times brighter than
its typical quiescence emission about one month prior to the burst.
In Fig.~\ref{AXP_DetSim},
we mark the elevated X-ray luminosities of SGR~1806$-$20
and SGR~1900$+$14 by stars.

Based on this plot we estimate that
the probability to detect a pulsating
X-ray source associated with an AXP/SGR
in M31, using the 50~ks {\it XMM}-Newton image we analyzed,
is $\sim10\%$ (per SGR/AXP).
We note however, that for a 2~Ms exposure using {\it XMM},
the probability to detect an AXP/SGR in
M31 increases to $\sim50\%$.

\section{Discussion}
\label{Disc}

In the following we discuss the
energetics, spectral and temporal properties of GRB~070201 (\S\ref{disc-en}).
Given the properties of this event we discuss its nature in \S\ref{disc-nat}.

\subsection{Energetics, spectrum and light curve}
\label{disc-en}

The IPN error quadrilateral of the bright
GRB\,070201 includes the outskirts of
the nearby (770~kpc)
galaxy M31.
If indeed GRB\,070201 has originated in M31,
the isotropic energy release from this burst,
$(1.41_{-0.18}^{+0.07})\times10^{45}$~erg,
is of the same order of magnitude as that emitted by SGR giant flares.
For comparison, the isotropic energy release of the 1979 March 5
SGR~0526$-$66 flare was $>6\times10^{44}\,$erg (Mazets et al. 1979), that
of the 1998 August 27 flare from SGR~1900$+$14
was $2\times10^{44}\,$erg
(Mazets et al. 1999),
while the energy release from the 2004 December 27
giant flare from SGR~1806$-$20 was as high as
$(1-4)\times10^{46}\,$erg
(Hurley et al. 2005; Palmer et al. 2005; Cameron et al. 2005).

In the context of the magnetar model for SGR giant flares
(Thompson \& Duncan 1995; 1996)
we expect the fireball to be optically
thick and therefore to produce a quasi-thermal spectrum.
%
As discussed in \S\ref{GRB070201},
the gamma-ray spectrum
of GRB~070201 (Golenetskii et al. 2007b), at peak luminosity,
as well as that of the SGR~1806$-$20
2004 Dec 27 giant flare (Frederiks et al. 2007a),
are not well described by a black-body spectrum.
However, this does not necessarily mean that
the spectrum of the burst is not a modified
thermal spectrum.
A simple consistency test for the SGR hypothesis
is to assume the spectrum is quasi-thermal;
we would then expect the black-body radius of the
emission region to be on the order of
the radius of a neutron star.
%
%
%
By approximating the gamma-ray spectrum of
SGR flares by a black-body spectrum,
one can derive a rough black-body radius for the
bursting source.
GRB~070201 had a peak luminosity
(on a 2~ms time scale) of $1.14_{-0.35}^{+0.20}\times10^{47}$~erg~s$^{-1}$,
and a peak energy of the observed
gamma-ray spectrum that corresponds
to black-body temperature of $\sim1.6\times10^{9}$~K.
Using the distance of M31,
we find a black-body radius of $60\pm40\,$km.
This radius is roughly consistent with
the sizes derived for other SGR giant flares
(e.g., Hurley et al. 2005; Ofek et al. 2006).

The temporal behavior of the gamma-ray emission from GRB~070201
(Fig.~\ref{GRB070201_LC}; see also Mazets et al. 2007)
is somewhat different from that of the
2004 December 27, SGR~1806$-$20 giant flare
(e.g., Hurley et al. 2005; Palmer et al. 2005; Terasawa et al. 2005).
In GRB~070201, the rise to maximum flux
is interrupted by two secondary peaks,
and the total rise time is somewhat longer
than in the case of the 2004 December 27 event.
Moreover, it seems that the light-curve of GRB~070201
is more variable than the typical SGR giant flare light curves.
Such variability is consistent with that seen in
the case of cosmological short-duration hard-spectrum GRBs
(e.g., Nakar \& Piran 2002; for a recent review see Nakar 2007).
However, our knowledge about SGR giant flare
light curves is based on a very small sample of events.

\subsection{The nature of GRB~070201}
\label{disc-nat}

Given the short-duration of this GRB
and its spatial association with M31,
there is a possibility that this burst
is an SGR flare in M31.
Estimating the probability for a chance coincidence
is susceptible to the pitfalls of {\it a-posteriori} statistics.
Keeping this in mind,
a rough estimate of the chance coincidence probability
is given by the sum of the area of M31 and the
error quadrilateral of GRB\,070201
(about $2$~deg$^{2}$), multiplied by the
number of short-hard GRBs detected by Konus-{\it Wind}
in the last 15\,yrs
($\sim30$; Ofek 2007a), and divided by the area of the
celestial sphere.
This rough chance coincidence probability is about $0.2\%$.
Therefore, we suggest that the simplest explanation
is that GRB\,070201 is indeed related to M31,
and that it was an SGR giant flare.
This is supported by the fact that, like other
known SGRs (Gaensler et al. 2001),
the GRB\,070201 error box is spatially associated
with star forming regions in M31 (Fig.~\ref{GALEX_X}).

We note that, if located in M31,
the energy of this event ($\sim10^{45}$~erg)
is too large for other kinds of known ``Galactic GRBs''
(e.g., Kasliwal et al. 2007).
Moreover, Abbott et al. (2007b) search for
a gravitational wave signal coincident
with the time of this burst using
the Laser Interferometer Gravitational wave Observatory
(LIGO).
The lack of signal argues against a compact
object merger (neutron stars/black-holes)
in M31, while is consistent with this event
being an SGR giant flare in the Andromeda galaxy.

Finally, we note that instruments like {\it Swift}-BAT
(Gehrels et al. 2004),
and {\it GLAST}-GBM (Band et al. 2004),
will be able to detect fainter bursts, with energy
of about $\sim10^{42}$~erg, from the Andromeda galaxy.
Such bursts are several orders of magnitude
more common than $\sim10^{45}$~erg events.
Therefore, with appropriate fast response
X-ray follow up observations of GRBs with
error regions that include nearby galaxies,
it may be possible to detect the afterglows
of such extragalactic SGR flares.

To summarize,
we do~not identify a visible light afterglow
associated with GRB\,070201.
Furthermore, we did~not find any periodic X-ray source
in archival {\it XMM} images of the intersection of
the error quadrilateral of GRB~070201 with M31.
We show
that the probability to detect a pulsating
X-ray source associated with an AXP/SGR
in M31, in the available {\it XMM} data,
is $\sim10\%$.
Therefore, the fact that we did not find
a X-ray pulsating source within the error quadrilateral
does not rule out the possibility that GRB~070201
is an SGR giant flare in M31.

\acknowledgments
This work is supported in part by grants from NSF and NASA.
HS acknowledges support from the Bundesministerium f\"ur Wirtschaft und 
Technologie/Deutsches Zentrum f\"ur  Luft- und Raumfahrt
(BMWI/DLR, FKZ 50 OR 0405).
MMK acknowledges the Moore Foundation for
the George Ellory Hale Fellowship.

\end{document}